\documentclass[prd,aps,epsf,amsfonts,floats,
amssymb,amsmath,nofootinbib]{revtex4}             

\usepackage{color}
\usepackage{graphicx}
\usepackage{amssymb}
\usepackage{hyperref}

\DeclareRobustCommand*\textsubscript[1]{%
  \@textsubscript{\selectfont#1}}
\def\@textsubscript#1{%
  {\m\ensuremath{_{\mbox{\fontsize\sf#1}}}}}

\newcommand{\be}{\begin{equation}}\newcommand{\ee}{\end{equation}}
\newcommand{\bea}{\begin{eqnarray}}\newcommand{\eea}{\end{eqnarray}}
\newcommand{\brr}{\begin{array}}\newcommand{\err}{\end{array}}
\newcommand{\bit}{\begin{itemize}}\newcommand{\eit}{\end{itemize}}
\newcommand{\ben}{\begin{enumerate}}\newcommand{\een}{\end{enumerate}}

\definecolor{darkred}{rgb}{.8,0,0}

\definecolor{darkblue}{rgb}{0,0,.8}

\def\lab{\label}

\def\de{\delta}

\def\1{{_{1}}}\def\2{{_{2}}}

\begin{document}

\title{Influence of gravity on the collective molecular dynamics of liquid water:\\ the case of the floating water bridge}

\author{Emilio Del Giudice${}^{1}$ 
and Giuseppe Vitiello${}^{2}$\footnote{Corresponding author. Email: vitiello@sa.infn.it}}

\vspace{2mm}


\address{${}^{1}$Istituto Nazionale di Fisica Nucleare, Sezione di Milano, Milano - 20133 Italy and\\ IIB, Neuss, Germany\\
${}^{2}$Dipartimento di Matematica e Informatica and\\
Istituto Nazionale di Fisica Nucleare, Gruppo collegato di Salerno, Universit\`a di Salerno, Fisciano (SA) - 84084 Italy
}

\begin{abstract}
Quantum electrodynamics (QED) produces a picture of liquid water as a mixture of a low density coherent phase and an high density non-coherent phase. Consequently, the Archimedes principle prescribes that, within a gravitational field,  liquid water should be made up, at surface, mainly of the coherent fraction, which becomes a cage where the gas-like non-coherent fraction is trapped, acquiring a non-vanishing pressure (vapor tension). Therefore, it is possible to probe the QED picture by observing the behavior of liquid water under reduced gravity conditions. The floating water bridge could be a useful test model.
\end{abstract}



\maketitle


The elucidation of internal structure of liquid water has attracted so far a large amount of work. Until recently the widespread opinion has considered liquid water as an array of molecules kept together by static interaction (H-bond, electric dipole-dipole interaction, and so on) \cite{Franks}. In recent time, however, a new approach has emerged where the interaction field was not only  a static one, but also the time-dependent electromagnetic (em) field. The problem of the interaction between atoms/molecules and the em field has been addressed within two different theoretical frameworks, QED \cite{Preparata,PRL88,Arani95,PRA2006} and quantum optics \cite{Kurcz1,Kurcz2}. Some widely used approximations in electrodynamics and optics, such as the slowly varying envelope approximation and the rotating wave approximation, have been dropped in these approaches. The emerging picture is the following:

\vspace{0.3cm}

i) above a density threshold and below a critical temperature the minimum energy state of the ensemble of molecules interacting with the em field is no longer the configuration where the oscillations (phases) of the molecules are uncorrelated and the em field is vanishing.
The new minimum energy state becomes a configuration where all molecules enclosed within an extended space region (coherence domain (CD)) get their phases locked in tune with a non-vanishing em field trapped within the CD.
The size of the CD is the wave length of this trapped em field. All component molecules oscillate in unison between their individual molecule ground state and an excited state, whose volume is of course wider than the ground state volume. The CD becomes a cavity for the em field because the dynamics gives to the photon an imaginary mass according to the well known Anderson-Higgs-Kibble mechanism~\cite{PRA2006}. The self-trapping of the em field guarantees that the energy of the CD has a finite lower bound.

ii) The above electrodynamic attraction is counteracted by thermal collisions which push molecules out of tune. Therefore, at non-vanishing temperature $T$, like in the Landau's picture of liquid Helium \cite{Landau}, the liquid becomes a two-phase system, where a fraction $F_{c}(T)$ behaves coherently, whereas a fraction $F_{nc}(T) = 1 - F_{c}(T)$ forms a dense gas trapped in the interstices among CDs. Since coherent molecules are wider than the non coherent ones, the density of the coherent fraction is lower than the density of the non-coherent fraction; in ref. \cite{Arani95} the density of the coherent fraction of water has been estimated to be $0.92$, the same value of the density of ice. According to the thermal dynamics in the bulk liquid there is a continuous crossover of molecules between the two fractions, so that, whereas the total number of coherent molecules remains constant at a given $T$, their space distribution changes continuously. This feature explains why experiments, like neutron scattering, having a resolution time longer than the typical time of flickering of space coherent structures,  probe the liquid as an homogeneous one. Only experiments having a resolution time short enough would detect the real inhomogeneous structure of liquid water \cite{PNAS}.

\vspace{0.3cm}

The above points characterize the bulk liquid. The situation changes at interfaces where the coherent structures are stabilized by the boundary conditions. The stabilization is provided near the walls by the attraction between the molecules of the liquid and the wall; at the air-liquid interface the surface liquid layer is almost completely coherent because of the difference of density between the two fractions.
Water CDs, because of the Archimedes principle, float and form a cage which keeps the gas-like non-coherent fraction trapped inside; the non-coherent fraction acquires consequently a pressure $P_{nc}(T)$, which gives rise to the surface tension. When temperature reaches a critical value $T_{b}$ such that $P_{nc}(T_{b})$ equals the external pressure the cage is broken, the non-coherent fraction flies away and the process of boiling starts. When $T < T_{b}$, only a number of non-coherent molecules, whose amount increases with $T$, is able to find its own way through the interstices among the CDs forming the surface layer and leave the liquid; this is the phenomenon of the evaporation.

The vapor-liquid transition occurs through the formation of coherence domains in all molecular species. The case of liquid water is peculiar \cite{Arani95} because in this case the molecule excited state involved in the coherent oscillation lies at an energy of $12.06 ~eV$ which is just below the ionization threshold at $12.60 eV$. The size of the water CD is the wavelength associated to the em mode at $12.06 ~eV$, namely $0.1 ~\mu m$.
Coherent molecules therefore oscillate between a state where all electrons  are tightly bound and a state where one electron is so loosely bound to be considered almost free. Consequently, whereas non-coherent water is an almost perfect insulator and chemically an oxidant, coherent water is a semiconductor and chemically a reducer; the interface coherent-water/non-coherent-water is therefore a redox pile and moreover a difference of electric potential could be found across it. In ref. \cite{Marchettini} this difference of potential has been estimated to be included in the interval between $50$ and $100 ~mV$. This difference of potential gives rise to voltage fluctuations in bulk water since the space distributions of the two fractions are flickering. On the contrary, at the surface of stabilized layers of coherent water, such differences of electric potential should become observable, as confirmed by the experiments performed on the interfacial water close to hydrophilic surfaces performed by the Pollack group \cite{Pollack}.
The existence of these differences of electric potentials plays a fundamental role in the dynamics of biological surfaces (cell membrane, and so on).

The existence of a reservoir of almost free electrons in the water CDs makes the CD as a whole an excitable object since the CD can accept small external supplies of energy producing coherent excitations (cold vortices) of the ensemble of almost free electrons. This phenomenon has been analyzed in ref. \cite{Boston}, where the spectrum of the excited state of the water CD has been derived. Since water CDs have acquired such non-trivial internal spectrum, there is the possibility of the onset of a coherence among them; in other words, water CDs could start to oscillate in unison in a region much more extended than the volume of the single CD. The spectrum of the water CD is extremely rich; the spacing among levels is in the order of radio-waves ($mm$-waves) and the upper limit is extremely high since is given \cite{Arani95} by the product of the energy gap of the water molecule ($0.26 ~eV$) times the number of the CD component molecules (some millions).
The excited state of the water CD cannot decay in a thermal way since in the cold vortices the motion of the quasi-free electrons is frictionless like in a superconductor (apart topological reasons).
The excited CDs could decay only by transferring their whole energy to some other resonating objects such as an ensemble of non-aqueous molecules able to oscillate with the same frequency. In this case, the water CD becomes an oscillator which picks up energy from the environment with high value of entropy, transforms it into a low entropy excitation energy of a coherent state and releases it as the chemical activation energy of some particular molecular species. As described elsewhere \cite{VoeikovDelGiudiceWaterJ}, water CDs should play an important role in the self-organization of a biochemical system. Moreover, as also remarked in ref. \cite{DelGiudFuchsVitielloWaterJ},  the presence of quasi-free electrons in the elementary CDs characterizes the dynamical regime of the system.  Since the motion of these electrons is confined within the CDs, it is necessarily a closed one, which implies that a magnetic field is thus generated.
Since the voltage $V$ is related to the phase $\phi$ through the relation
\bea \lab{pot}
V = - \frac{\hbar}{e} \frac{d \phi}{dt}~,
\eea
%
where $e$ is the electron charge and $\hbar$ is the Planck constant divided by $2\pi$, the application of a voltage implies a variation of the phase $\phi$ which adds up to the original phase of the unperturbed CDs. Should this additional  phase generated by the voltage be dominant with respect to the original phase (as in the case of high  applied potential), the new phase involves a macroscopic region and is thus space-correlating all the phases of the CDs enclosed in such a region. A coherence among the CDs then may emerge \cite{DelGiudTedeschi}. In this new macroscopic coherent region a definite non-vanishing gradient of the phase thus appears, that in turn produces a non-vanishing magnetic potential according to the relation
\bea \lab{pot}
{\bf A} = \frac{\hbar c}{e}~ {\bf grad}\,\phi ~.
\eea
%
As a final result, coherence is then established on  a scale much larger than the original $0.1 ~\mu m$.

The increase of the size of the coherent region makes possible the appearance of a new phenomenon: levitation of droplets of water. We have said above that water CDs enclose an ensemble of almost free electrons. In ref. \cite{Arani95} it has been shown that the statistical weight of the excited water molecule state, where one electron is almost free, in the coherent superposition of the ground and the excited state is about $0.13$. This means that the ensemble of almost free electrons includes $0.13$ electrons per water molecule. It is a well known result \cite{Feynman} that an ensemble of coherent electrons, like for instance in a superconducting metal, is able to expel the magnetic field from its interior apart a boundary region whose depth is the so-called penetration length $l_M$ (Meissner effect). The Meissner penetration length $l_M$ is estimated to be (see Eq. (21.25) in ref. \cite{Feynman})
\bea \lab{Ml}
l_M = \frac{1}{\sqrt{4\pi n N r_{0}}} ~,
\eea
where $r_{0}$ is the 'electromagnetic radius" of the electron ($r_{0} = 2.8 \times 10^{-13} ~cm$), $N$ is the number of electrons per $cm^3$ and $n$  could be $2$ or $1$, according to the fact that electrons are Cooper-paired or not. In the case of a superconducting metal having a density of $3 \times 10^{22}$ atoms per $cm^3$ with one conduction electron per atom, where electrons are Cooper-paired, one gets $l_M \approx 0.2 ~\mu m$. In the case of a water CD at room temperature, where the density is $3 \times 10^{22}$ molecules per $cm^3$, the Cooper pairing is absent and each molecule contribute  $0.13$ electrons, one gets
\bea \lab{Ml2}
l_M \approx 0.8 ~\mu m ~.
\eea
Since a water CD has a size of $0.1 ~\mu m$ \cite{Arani95}, the Meissner effect should be absent in normal liquid water, in agreement with observed facts: the magnetic field is not expelled from water. However, the situation could be entirely different when coherence among neighbouring CDs appears, since the coherent region could then acquire a size larger or much larger than the value given by Eq.~(\ref{Ml2}). In this case the possibility of a magnetic levitation would appear. Let us consider a droplet of coherent water having a size $R$, which is floating on the water surface. A strong vorticity is maintained in the surface layer of water by a strong electric potential. Consequently, a large magnetic field $H$ is produced, which is present in the outer regions of CDs; this field decreases sharply in the surroundings so that the water droplet on the surface is immersed in a strongly inhomogeneous field, having a high uniform value below the droplet and a much lower value above it; we could just neglect the field above the droplet, and assume it as vanishing. Moreover the magnetic permeability $\mu$ is zero within the coherent region and is approximately one in the outer region, whose depth is $l_M$. Therefore, by calling $d$ the depth of the surface layer, we could approximate $\mu\,{\bf grad}\,H^2$ as $\frac{\mu\,H^2}{d}$ and $H^2 {\bf grad}\,\mu$ as $\frac{H^2 \mu}{l_M}$. Since $d \gg l_M$ we can neglect the upward force $-\mu\,{\bf grad}\,H^2$ with respect to $-H^2\,{\bf grad}\,\mu$. The energy density of the magnetic field is:
\bea \lab{ende}
U = \frac{1}{2}~\mu\, H^2 ,
\eea
which gives rise to a force per unit volume:
\bea \lab{force}
{\bf F} = - {\bf grad}\,U = - \frac{1}{2}\, H^2 \,{\bf grad}\,\mu - \frac{1}{2}\,\mu  \,{\bf grad}\,H^2 \approx - \frac{1}{2}\, H^2 \,{\bf grad}\,\mu
\eea
acting on the droplet. The total force ${\bf F}_L$ acting on the droplet is the resultant of an upward force acting on the surface of the lowest region of the droplet and a downward force acting on the top region. However, this last term is negligible since the value of $H^2$ above the droplet is very small, so that we have only the upward term. By calling $A$ a numerical constant which depends on the geometry of the droplet and whose order of magnitude is $O(1)$, $S$ the cross-section surface of the droplet and ${\bf u}$ a unit vector pointing upward,  ${\bf F}_L$ is
\bea \lab{forcelev}
{\bf F}_L= {\bf F} V = A\,S\,l_M \,{\bf F} = A\,S\,H^2  \mu \,{\bf u}~.
\eea
The intensity of this force should be compared to the weight $P$ of the droplet in air:
\bea \lab{Pdrop}
P = \rho_{coherent}\, g\, V  = B \rho_{coherent}\, g\, S\,R~,
\eea
where $B$ is a numerical constant whose order of magnitude is $O(1)$, $g$ is the gravity acceleration and $\rho_{coherent}$ the density of the coherent water. From Eq.~(\ref{forcelev}) and Eq.~(\ref{Pdrop}), by assuming that $\mu$ can be taken to be one, we get the condition for levitation:
\bea \lab{levit}
H^2 >  C \,\rho_{coherent}\, g\, R ~,
\eea
where $C$ is the numerical constant $\frac{B}{A}$, whose order of magnitude is of course $O(1)$.

So, when the vorticity becomes strong enough to produce a magnetic field able to meet the condition (\ref{levit}), levitation would appear, provided that  $R \gg l_M = 0.8 ~\mu m$.

\vspace{.3cm}

The prediction of such effect can be verified and give a check of the theory. Recently, an experiment has been produced \cite{Fuchs07} that allows to perform such a check, namely the formation of the floating water bridge \cite{Armstrong93}. Two beakers filled with pure water up to a small distance $\de$ from the rim are placed near by. Two electrodes connected with a voltage generator are immersed one in each beaker and a voltage of several thousands of Volts (typically between $15000$ and $30000$ $V$) is established. In a matter of few seconds vortices appear (mainly in the beaker with the positive electrode), subsequently columns of water, some millimeter thick arise from the water surface along the inner walls of the beaker and an arch is produced where water self-pipes itself and is able to bridge the distance between the glasses when we bring them as far as a few centimeters. A number of interesting properties of the bridge have been also detected and described \cite{Fuchs07,Fuchs08,Fuchs09,Nishiumi09,Woisetschläger09,Fuchs10,Widom09,24,25}. We have interpreted elsewhere \cite{DelGiudFuchsVitielloWaterJ} the formation and the properties of the floating water bridge in the theoretical frame depicted above. The main steps of this interpretation are the following:

1. The surface layer of the water in the beakers is made up mainly of coherent water. The different CDs are still not correlated in phase.

2. The application of a very high potential produces a phase agreement among these coherent domains which gives rise to macroscopic vortices \cite{DelGiudFuchsVitielloWaterJ}. These produce in turn an extended coherence.

3. When the critical Meissner threshold is overcome the magnetic levitation force appears and upward water flows arise \cite{Fuchs07,Woisetschläger09}; these flows involve coherent water only, because non-coherent water is not affected by the Meissner effect.

4. The em fields trapped  in the CDs give rise to exponentially falling tails protruding from the CDs (evanescent fields) on a scale of their wavelengths, namely in the order of a fraction of a $\mu m$, which is larger than the average distance among CD (droplets). A long range attractive force develops among them and gives rise to the formation of the arch (the floating bridge).

5. The existence of the coherence among the CDs forming the arch stabilizes it.

6. The electric field which is the consequence of the high voltage acts on the CDs voltage (for a detailed description of the electric field see ref.  \cite{Widom09}). On the boundary of the CDs the presence of a gradient of electric potential induces the formation of a double layer of charges \cite{Marchettini}; the external layer is negatively charged. Under the action of the electric field the bulk of the CDs are pushed in the direction from the positive to the negative potential; whereas the negatively charged outer layer slides along the bridge in the opposite direction.

7. The simultaneous presence of magnetic field produced by vorticity makes this motion helicoidal. This has been shown experimentally \cite{Woisetschläger09} and explained theoretically \cite{DelGiudFuchsVitielloWaterJ}.


The above sequence of events is made possible by the presence on the surface of the beakers of an almost completely coherent layer of water, which in turn depends on the difference of density between coherent and non-coherent fractions. In other words, the  necessary preconditions for the formation and existence of the floating water bridge depend on the Archimedes force per unit volume $F_A = g(\rho_{average} - \rho_{coherent})$, where $\rho$ is the density,  since this force is responsible for the existence of the coherent layer on top of the water. In a system devoid of gravitation, such a layer would not form, and the CDs would be evenly distributed throughout the whole water volume. It is thus apparent that, should $g = 0$, this force would disappear and the formation of the floating water bridge would be impossible in any physical condition. Obviously, the "exclusion zone" top layer of the water described in Ref. \cite{Pollack} would not form in these conditions either, since it resembles the coherent water fraction floating above the non-coherent one.

If gravity were to be slowly removed from an existing water bridge, we predict the following scenario: If $g$ is reduced but not cancelled, the condition (\ref{levit}) for the threshold value of $H^2$ is eased since the threshold value, which is proportional to $g$, is decreased, allowing levitation at smaller values of the vorticity and enhancing the upward motion  of the droplets at a fixed value of the vorticity. Consequently, a decrease of gravity force should produce at first a more robust phenomenon. However, the further decrease of $g$, while still easing the constraint given by the inequality (\ref{levit}), eventually undermines the whole phenomenon since destroys its necessary precondition. At vanishing $g$ the Archimedes force which creates a wholly coherent surface layer disappears and the dynamics at root of its formation and existence would fade away.

In conclusion we would predict that the bridge's stability and probably diameter would increase, since in reduced gravity conditions the Archimedes force is still active, and gravity which is counteracting the electric and magnetic levitation forces would be reduced, resulting in a larger amount of water being drawn into the bridge. However, at a certain point, the turbulence caused by the water flow would push more and more of the CDs down into the bulk which would take a longer time to come up again due to the reduced Archimedes force. A precise calculation of this point comprises the evaluation of all acting forces - fluid dynamics, magnetic, electric and gravitational - and is clearly beyond the scope of this work. We plan to address this issue in more detail in the future. In this work, we restrain ourselves to the purely qualitative statement that at a certain point during the reduction of gravity, after an initial stabilization the floating water bridge would undoubtedly collapse. Likewise, the exclusion zone water top layer would disappear due to thermic fluctuations, albeit probably at a lower level of gravitation than that required to destroy the water bridge.

These predictions are a rigorous consequence of the QED picture of liquid water and we suggest performing experiments in order to prove or disprove their validity.


The authors wish to thank Prof. Elmar C. Fuchs for illuminating discussions on the phenomenic features of the floating water bridge.
Partial financial support from University of Salerno and Istituto Nazionale di Fisica Nucleare is also acknowledged.


\begin{thebibliography}{99}


\bibitem{Franks} Franks F (Ed), {\it A comprehensive treatise}, Plenum, N.Y. 1972-1982

\bibitem{Preparata} G. Preparata, {\it QED coherence in matter}, World Scientific 1995

\bibitem{Arani95} Arani R, Bono I, Del Giudice E, Preparata G, Int. J. Mod.  Phys. B 9 (1995) 1813-1841

\bibitem{PRL88} Del Giudice E, Preparata G, Vitiello G, Phys. Rev. Lett. 61 (1988) 1085-1088

\bibitem{PRA2006} Del Giudice E,  Vitiello G, Phys. Rev. A 74 (2006) 022105-9

\bibitem{Kurcz1} Kurcz A, Capolupo A, Beige A,  Del Giudice E, Vitiello G, Phys. Rev. A  81 (2010) 063821

\bibitem{Kurcz2} Kurcz A, Beige A, Capolupo A, Del Giudice E, Vitiello G, Phys. Lett. A  374 (2010) 3726-3732

\bibitem{Landau} Landau L D, J. Physics USSR (Moscow) 5 (1941) 71

\bibitem{PNAS} Huan C, et al., PNAS Proc Natl Acad Sci. USA 106 (2009) 15241-6

\bibitem{Marchettini} Marchettini N, Del Giudice E, Voeikov V, Tiezzi E, J. Theor. Bio. 265 (2010) 511-516

\bibitem{Pollack} Zheng, J.-M.; Chin, W.-C.; Khijniak, E.; Khijniak, E., Jr.; Pollack, G.H.
Ad. Colloid. Interface Sci. 23 (2006) 19-27

\bibitem{Boston} Del Giudice E, Preparata G,
in {\it Macroscopic quantum coherence}, Sassaroli et al. Eds., Worl Sc., Singapore (1998)

\bibitem{VoeikovDelGiudiceWaterJ} Voeikov V, Del Giudice E, Water J. 1 (2009) 52-75 

\bibitem{DelGiudFuchsVitielloWaterJ} Del Giudice E, Fuchs E, Vitiello G, Water J. 2 (2010) 69-82

\bibitem{DelGiudTedeschi} Del Giudice E, Spinetti P R, Tedeschi A,  Water 2 (2010) 566-586

\bibitem{Feynman} Feynman R P, Leighton R B, Sands M, {\it The Feynman lectures on Physics}, Addison-Wesley Publishing Co., N.Y. (1965), Vol. 3, Section 21-6

\bibitem{Fuchs07} Fuchs E C, Woisetschl\"ager J, Gatterer K, Maier E, Pecnik R,
Holler G and Eisenk\"olbl H, J. Phys. D: Appl. Phys. 40 (2007)
6112-6114

\bibitem{Armstrong93} Armstrong W G, {\it Electrical phenomena The Newcastle Literary and
Philosophical Society, The Electrical Engineer} (1893) 10 February
1893, pp 154-145

\bibitem{Fuchs08} Fuchs E C, Gatterer K, Holler G and Woisetschl\"ager J 2008, J.
Phys. D: Appl. Phys. 41 (2008) 185502-7

\bibitem{Fuchs09} Fuchs E C, Bitschnau B, Woisetschl\"ager J, Maier E, Beuneu B, Teixeira J, J. Phys. D: Appl. Phys. 42 (2009) 065502-6


\bibitem{Nishiumi09}Nishiumi H, Honda F, Res. Let. Phys. Chem. 2009 (2009)  art. ID 371650-3 


\bibitem{Woisetschläger09} Woisetschl\"ager J, Gatterer K, Fuchs E C, Exp Fluids (2010) 48:121–131

\bibitem{Fuchs10} Fuchs E C, Baroni P, Bitschnau B, Noirez L, J. Phys. D: Appl. Phys. 43 (2010) 105502-5 

\bibitem{Widom09} Widom A, Swain J,  Silverberg J,  Sivasubramanian S, Srivastava Y N, Phys. Rev. E 80 (2009) 016301-7 


\bibitem{24} R. C. Ponterio, M. Pochylski, F. Aliotta, C. Vasi, M. E. Fontanella, F. Saija, J. Phys. D: Appl. Phys. 43 (2010) 175405-8 

\bibitem{25} Fuchs E C, Water 
 2 (2010) 381-410

\end{thebibliography}

\end{document}